\documentclass[a4paper, 11pt]{article} 

\usepackage{geometry} 
\usepackage{graphicx} 

\usepackage{float} 
\usepackage{wrapfig} 

\usepackage{lipsum} 

\usepackage{amsmath} 

\usepackage{undertilde}

\usepackage{multirow}

\usepackage{indentfirst}

\usepackage{enumerate}

\graphicspath{{Pictures/}} 

\title{\textbf{Non-Gaussian Stochastic Volatility Model with Jumps via Gibbs Sampler}} 

\author{\textsc{Arthur T. Rego and Thiago R. dos Santos} 
	\\{\textit{Universidade Federal de Minas Gerais, Brazil}}} 
\date{\vspace{-5ex}}

\begin{document}
	
	\maketitle 
	
	
	
	\begin{abstract}
			In this work we propose a model for estimating volatility from financial time series, extending the non-Gaussian family of space-state models with exact marginal likelihood proposed by Gamerman, Santos and Franco (2013). On the literature there are models focused on estimating financial assets risk, however, most of them rely on MCMC methods based on Metropolis algorithms, since full conditional posterior distributions are not known. We present an alternative model capable of estimating the volatility, in an automatic way, since all full conditional posterior distributions are known, and it is possible to obtain an exact sample of volatility parameters via Gibbs Sampler. The incorporation of jumps in returns allows the model to capture speculative movements of the data, so that their influence does not propagate to volatility. We evaluate the performance of the algorithm using synthetic and real data time series and the results are satisfactory.

	\end{abstract}
	
	\hspace*{3,6mm}\textit{Keywords:} Financial time series, stochastic volatility, Gibbs Sampler, dynamic linear models. 
	
	\vspace{30pt} 
	

\section{Introduction} 

Understanding the behavior of asset prices is essential for capital allocation decisions between the available investment option. Such decision depends on what one thinks about risks and returns associated with these investment options. The most accepted theory is that the returns on high volatility assets follow a random walk with some outliers points, that usually occur during abnormal volatility increases, such as in financial and political crisis events. The future returns would be unpredictable, but the volatility can be estimated and monitored in order to detect such events and anticipate their movements. Under the Bayesian perspective, the inferential procedure of the stochastic volatility models commonly used are mostly based on intensive computational methods, e.g., Markov Chain Monte Carlo (MCMC) methods using Metropolis-Hastings algorithms, which raises questions about the usage of more automatic and simpler computational implementation methods that can be used to bring fast and reliable results. Dealing with financial time series bring three main challenges. They include finding a model that: fits well to the data, accommodates the heavy tails that exists in non-Gaussian returns; is fast enough to bring results on time to be used by market agents; and is flexible to include new source of data and accommodate outliers and skewness, that improve the model fit. 

Many models have been developed for risk measuring purposes. For example, Eraker, Johannes e Polson (2003) adopt the Stochastic Volatility model (SV) and study the influence of inserting jumps to improve the model. They suggest including jumps on returns and volatility in order to improve the model dynamic in case of spot changes in volatility, as in financial crisis moments. Omori \textit{et. al} (2006) include the leverage effect on SV models, which refers to the increase in volatility following a previous drop in stock returns, and model it by the negative correlation coefficient between error terms of stock returns. Nakajima and Omori (2007) extend its application to SVJ models with heavy-tail distribution, obtained by a scale mixture of a generalized gamma distributed mixture component together with a normal distributed error in order to generate generalized Skew-t distributed innovations, discussing the fit gains on including such feature. Warty, Lopes and Polson (2017) investigate sequential, or online, Bayesian estimation for inference of stochastic volatility with variance-gamma jumps in returns and compare its performance to the model that uses offline Markov Chain Monte Carlo. Kanaya and Kristensen (2016) use a stochastic volatility model with a two-step estimation method. In the first
step, they nonparametrically estimate the instantaneous volatility process, and in the second step, standard estimation methods for fully observed diffusion processes are employed, but with the filtered or estimated volatility process replacing the latent process. Bandi and Reno (2018) also provide nonparametric methods for stochastic volatility modeling, allowing the joint evaluation of return and volatility dynamics with nonlinear drift and diffusion functions, nonlinear leverage effects, and jumps in returns and volatility with state-dependent jump intensities. However, those models are either nonparametric or the parameter estimation is somewhat complex, since there are no closed-form to the full conditional posterior distributions, being necessary the use of MCMC methods with Metropolis Hastings (MH) steps.

The main aim of this work is to find an alternative model that accommodates speculative financial asset returns data, allowing the innovations to assume heavy tails distributions, includes jumps on returns in order to get the impact of uncommon events on financial markets,and has a simpler and more automatic inferential procedure, like Gibbs Sampling together with a block sampling structure, for estimating the model parameters, mitigating convergence issues. We introduce a non-Gaussian stochastic volatility model with jumps, as well as an application to real S$\&$P-500 index and Brent Crude future returns time series.

\section{The non-Gaussian stochastic volatility model with jumps on returns} 

The non-Gaussian stochastic volatility model with jumps on returns (NGSVJ) for time series $\left\lbrace y_t \right\rbrace_{t=1}^n$ is given by:

\begin{eqnarray}	
y_{t}&=& \mu + J_{t}^{y} + \upsilon_{t}, \quad where \quad \upsilon_{t}|\gamma_{t} \sim N(0,\gamma_{t}^{-1}\lambda_{t}^{-1}) ,\\
\lambda_{t} &=& \omega^{-1}\lambda_{t-1}\zeta_{t}, \quad where \quad \zeta_{t}|Y_{t-1},\varphi \sim Beta(\omega a_{t-1},(1-\omega)a_{t-1}), \mbox{ where} 
\end{eqnarray}
$$J_{t}^{y}=\xi_{t+1}^{y}N_{t+1}^{y}
, \quad
\xi^{y}\sim N(\mu_{y},\sigma_{y}^{2}) , \quad and \quad
Pr(N_{t+1}^{y}=1)=\rho_{y}
.
$$

In this model, $y_{t}$ represents the log-return in percentage, defined as $y_{t} = 100*(log(S_{t})-log(S_{t-1}))$, where $S_{t}$ is the asset price on time t. $J_{t}^{y}$ is the jump, composed by the jump indicator $N_{t+1}^{y}$ and magnitude $\xi^{y}\sim N(\mu_{y},\sigma_{y}^{2})$, in the same way proposed by Eraker \textit{et al.} (2003). $\mu$ represents the equilibrium log-return of $y_{t}$.

$\gamma_{t}$ is the variance mixture component [Gamerman \textit{et al.}, 2013].  Using
$\gamma_{t}\sim\Gamma(\frac{\nu}{2},\frac{\nu}{2})$, the unconditional distribution of errors assume  a $t_{\nu}(0,1)$ distribution. $\lambda_{t}^{-1}$ is the volatility of returns, and the main interests lays on estimating its value over time, since it is the main variable on risk and stock options pricing.  $\omega$ is a discount factor and is specified, in order to avoid a MH step for its estimation, keeping the estimation procedure more automatic via Gibbs Sampling. $a_{t-1}$ is the shape parameter of the filtering distribution of $\lambda_{t}$, which is described on details in Gamerman \textit{et al.} (2013).


The model provides the needed flexibility by using the SSM form, with mixtures on variance in order to achieve non-Gaussian distribution for innovations. It has also a formulation that allows the full conditional posterior distributions to be available, so that Gibbs Sampler can be used to sample from the conditional posterior distributions, bringing implementation simplicity to the model.

In this case, there are no dimensionality issues with the parametric space, since all full conditional posterior distributions are obtained through the model properties and can be sampled via Gibbs Sampler. Using proper priors to parameters it is possible to obtain the full conditional posterior distribution, where the priors are chosen in order to obtain a conjugate posterior distribution. Another advantage lays on sampling mean $\mu$ and volatility $\lambda_{0:n}$ in blocks, speeding up the sampling process. A more detailed description for the model procedures can be seen in Gamerman \textit{et al.} (2013).

The procedure of including jumps, adapted from Eraker \textit{et al.} (2003), gets advantage of models properties that guarantee a simpler computational implementation sampling method. For this work the concern is about including jumps only in returns, since including them also on volatility requires a more complex structure in order to preserve the model properties and a fast inference procedure, and will be addressed to future works.

\subsection{Bayesian Inference} %

For mean parameter of the log-returns , $\mu$, a prior N($m_{0}$,$C_{0}$) is specified and the samples of its posterior distribution can be obtained through a standard FFBS algorithm.

For ease notation, let $\Phi = (\mu,\utilde{J^{y}}, \utilde{\gamma}, \utilde{\lambda},\utilde{\mu_{y}},\utilde{\sigma_{y}^{2}},\utilde{\xi},\utilde{\rho},\utilde{N^{y}})$, excluding the parameter being evaluated, i.e. $\Phi_{[-\lambda]} = (\mu,\utilde{J^{y}}, \utilde{\gamma},\utilde{\mu_{y}},\utilde{\sigma_{y}^{2}},\utilde{\xi},\utilde{\rho},\utilde{N^{y}})$ .

With the prior distribution for $\lambda_{t}$, that is given by $\lambda_{t}|Y_{t-1}\sim Gamma(\omega a_{t-1}, \omega b_{t-1})$, and following the method proposed by Gamerman \textit{et al.} (2013), the updating distribution is:
\begin{equation}
p(\lambda_{t}|Y_{t},\Phi_{[-\lambda]}) \sim \text{Gamma}\left(\omega a_{t-1}+\frac{1}{2}, \omega b_{t-1}+\gamma_{t}\frac{(y_{t}-\mu-J_{t})^{2}}{2}\right).
\end{equation}
 
The procedure for sampling from $(\lambda_{t}|Y_t, \Phi_{[-\lambda]})$ can be seen on Appendix A [Gamerman \textit{et al.}, 2013]. $\omega$ is a fixed discount factor.

For the mixture component $\gamma_{t}$, a prior Gamma($\frac{\nu}{2},\frac{\nu}{2}$) is defined, which, when mixed as $\gamma_{t}^{-1}$, resulting in Inverse-Gamma, leads to a Student-t with $\nu$ degrees of freedom to the innovations. The full conditional posterior distribution is:
\begin{equation}
 p(\gamma_{t}|Y_{t},\Phi_{[-\gamma]})  \sim \text{Gamma}\left(\frac{\nu}{2}+\frac{1}{2}, \frac{\nu}{2}+\lambda_{t}\frac{(y_{t}-\mu-J_{t})^{2}}{2}\right).
\end{equation}

The parameter $\nu$ will be specified, since its posterior distribution does not have closed form, leading to a Metropolis step. A sensibility analysis is made for it, comparing models selection criterion statistics for different values of $\nu$. Recall that the main objective of the NGSV model is keeping an automatic and simple procedure for the model estimation and, once it is determined for a specific asset time series, $\nu$ does not need to be changed for future observations.

The jump sizes $\xi_{t+1}^{y}$ follow a N($\mu_{y}$,$\sigma_{y}^{2}$). For the mean $\mu_{y}$ a non-informative prior N(m,v) is set, resulting in a full conditional posterior: 
\begin{equation}
p(\mu_{y}|Y_{n},\Phi_{[-\mu_{y}]}) \sim \text{N}\left(\frac{m\sigma^{2}_{y}+vn\bar{\xi^{y}}}{\sigma^{2}_{y}+nv},\frac{v\sigma^{2}_{y}}{\sigma^{2}_{y}+nv}\right).
\end{equation}

For the variance $\sigma_{y}^{2}$ a prior InverseGamma($\alpha,\beta$) is assumed, resulting in the full conditional posterior:
\begin{equation}
 p(\sigma^{2}_{y}|Y_{n},\Phi_{[-\sigma^{2}_{y}]}) \sim InverseGamma\left(\alpha+\frac{n}{2},\beta+\frac{\sum_{\substack{i=1 \\ J_{i}\neq 0}}^{t}(\xi_{i+1}^{y}-\mu_{y})^{2}}{2}\right).
\end{equation}
 
In both cases, n is the number of times that the jump is observed, and $\bar{\xi^{y}}$ the mean of jump sizes $\xi^{y}$. As the prior of jump sizes is assumed to be Normal, the full conditional posterior is also Normal, given by:
\begin{equation}
p(\xi_{t+1}^{y}|Y_{t},\Phi_{[-\xi]})\sim N\left(\frac{\mu_{y}\gamma_{t}^{-1}\lambda_{t}^{-1}+y_{t}\sigma^{2}_{y}-\mu\sigma^{2}_{y}}{\sigma^{2}_{y}+\gamma_{t}^{-1}\lambda_{t}^{-1}},\frac{\sigma^{2}_{y}\gamma_{t}^{-1}\lambda_{t}^{-1}}{\sigma^{2}_{y}+\gamma_{t}^{-1}\lambda_{t}^{-1}}\right).
\end{equation}

For jump probabilities $\rho$, a prior $Beta(\alpha,\beta)$ is set. The full conditional posterior is given by:
\begin{equation}
p(\rho|Y_{n},\Phi_{[-\rho]}) \sim Beta\left(\alpha+\sum_{i=0}^{n}N^{y}_{i},\beta+n - \sum_{i=0}^{n}N^{y}_{i}\right).
\end{equation}

Since the jump indicator $N^{y}$ can assume only two values, 0 or 1. The probability of observation at $t+1$ be a jump is given by:
\begin{equation}
P(N^{y}_{t+1}=1|Y_{t+1},\Phi_{[-N]}) \propto \rho P(Y_{t+1}|N^{y}_{t+1}=1,\Phi_{[-N]}).
\end{equation}
which is easy to calculate, since $P(Y_{t+1}|N^{y}_{t+1}=1,\Phi_{[-N]})$ is a Normal distribution. Using the concept proposed by Brooks and Prokopczuk (2011), if $P(N^{y}_{t+1}=1|Y_{t+1},\Phi_{[-N]})$ is greater then a threshold $\alpha$, then $N^{y}_{t+1}=1$. The threshold $\alpha$ in chosen such that the number of jumps identified corresponds to the estimate of the jump intensity $\rho$.

\subsection{Gibbs Sampler}

Let $Y_{n} = \{y_{t}\}^{n}_{t=1}$, $\utilde{J} = \{J^{y}_{t}\}^{n}_{t=1} =  \{\xi^{y}_{t+1}N^{y}_{t+1}\}^{n}_{t=1}$, $\utilde{\gamma} = \{\gamma_{t}\}^{n}_{t=1}$, $\utilde{\lambda} = \{\lambda_{t}\}^{n}_{t=1}$, $\utilde{\xi} = \{\xi^{y}_{t+1}\}^{n}_{t=1}$, $\utilde{N} = \{N^{y}_{t+1}\}^{n}_{t=1}$ and prior probability density $\pi(\gamma), \pi(\mu_{y}), \pi(\sigma^{2}_{y}), \pi(\xi), \pi(\rho_{y}) $ are set for $\utilde{\gamma}, \mu_{y}, \sigma^{2}_{y}, \utilde{\xi}, \rho_{y} $. Then, a sample of size M from the joint posterior distribution $\pi(\mu, \utilde{\lambda}, \utilde{\gamma}, \mu_{y}, \sigma^{2}_{y},\utilde{J},\rho_{y}|Y_{n})$ is drawn via Gibbs Sampler, whose follows:

	\begin{enumerate}[i]
		\item Initialize $\mu^{(0)}, \utilde{\lambda}^{(0)}, \utilde{\gamma}^{(0)},  \mu_{y}^{(0)}, (\sigma^{2}_{y})^{(0)}, \utilde{\xi}^{(0)}, \utilde{N}^{(0)} \text{ and } \rho_{y}^{(0)} $.
		\item Set $j=1$.
		\item Sample $\mu^{(j)}|Y_{n},\utilde{J}^{(j-1)},\utilde{\lambda}^{(j-1)},\utilde{\gamma}^{(j-1)}$ using FFBS algorithm.
		\item Block sample $\utilde{\lambda}^{(j)}|Y_{n},\mu^{(j)},\utilde{J}^{(j-1)},\utilde{\gamma}^{(j-1)}$ using algorithm described on Appendix A.
		\item Block sample $\utilde{\gamma}^{(j)}|Y_{n},\mu^{(j)},\utilde{J}^{(j-1)},\utilde{\lambda}^{(j)}$ as in  Eq.(4).
		\item Sample $\mu_{y}^{(j)}|\utilde{\xi}^{(j-1)},(\sigma^{2}_{y})^{(j-1)}$ as in  Eq.(5).
		\item Sample $(\sigma^{2}_{y})^{(j)}|\utilde{\xi}^{(j-1)},\mu_{y}^{(j)}$ as in  Eq.(6).
		\item Block sample $\utilde{J}^{(j)}|Y_{n},\mu^{(j)},\utilde{\lambda}^{(j)},\utilde{\gamma}^{(j)},\mu_{y}^{(j)},(\sigma^{2}_{y})^{(j)}$ by
		\subitem Block sample $\utilde{\xi}^{(j)}|Y_{n},\mu^{(j)},\utilde{\lambda}^{(j)},\utilde{\gamma}^{(j)},\mu_{y}^{(j)},(\sigma^{2}_{y})^{(j)}$ as in  Eq.(7).
		\subitem Block sample $\utilde{N}^{(j)}|Y_{n},\mu^{(j)},\utilde{\lambda}^{(j)},\utilde{\gamma}^{(j)},\utilde{\xi}^{(j)}$ as in Eq.(9).
		\item Sample $\rho_{y}^{(j)}|\utilde{J}^{(j)}$ as in  Eq.(8).
		\item Set $j = j+1$.
		\item If $j\leq M$, go to 3, otherwise stop.
	\end{enumerate}

Since all full conditional posterior distribution have closed-form, only Gibbs Sampler steps are used.

\subsection{Model Diagnostics and Specification Tests}

The approach to compare different specifications for model parameters is the BIC and DIC criteria, defined by:
\begin{eqnarray}
BIC &=& -2\,ln(\hat{L}) +k\times ln(n),\\
DIC &=& D(\utilde{y},\bar{\Phi}) + 2 pD,\\
pD &=& \bar{D}(\utilde{y},\Phi) - D(\utilde{y},\bar{\Phi}).
\end{eqnarray}
Where, for BIC, $\hat{L}$ is the maximized value of the likelihood function for the model, $k$ the number of parameters evaluated and $n$ the sample size, and, for DIC, statistical deviance $D(y,\Phi)$ is defined as:
\begin{eqnarray}
D(\utilde{y},\Phi) &=& -2 \, ln \left( p(\utilde{y}|\Phi) \right)
\end{eqnarray}
for data y and model parameters $\Phi$. The posterior mean deviance is given by:
\begin{eqnarray}
\bar{D}(\utilde{y},\Phi) = E \left[ D(\utilde{y},\Phi) | \utilde{y} \right].
\end{eqnarray}

\section{Simulation}

To illustrate the performance of the NGSVJ, we apply the method to synthetic data from the model proposed by  Warty et al (2017). To generate the volatility, we use:
\begin{equation}
v_t = v_{t-1} + \kappa(\theta - v_{t-1})\Delta +\rho\sigma_v\sqrt{v_{t-1}\Delta}\epsilon_{1,t}+\sigma_v\sqrt{(1-\rho^2)v_{t-1}\Delta}\epsilon_{2,t}
\end{equation}
where $\epsilon_{1,t}$ and  $\epsilon_{2,t}$ $\sim N(0,1)$. Synthetic data for returns is then generated from:
\begin{equation}
r_t = N(\mu + \mathcal{J}_t,\gamma_{t}^{-1}v_t), \qquad \mathcal{J}_t=N_{t}\xi_{t},
\end{equation}
where the jump times, $N_{t}$ are generated from a Bernoulli($\rho_y$), jump sizes $\xi_{t}$ from  $N(\mu_y,\sigma_y^2)$, and $\gamma_t$ from Gamma$(\frac{\nu}{2},\frac{\nu}{2})$. Setup of parameters was: log-returns mean $\mu = 0.05$; jump probability $\rho_y = 0.015$; jump magnitude mean $\mu_y=-2.5$ and standard deviation $\sigma_y=4$; variance mixture degrees of freedom $\nu=30$; volatility components $\Delta=1$, $\theta = 0.8$, $\kappa=0.015$, $\sigma_v=0.1$, $\rho=0.4$, same used by Warty et al (2017).

\begin{figure}[H]
	\centering
	\includegraphics[width=1\linewidth]{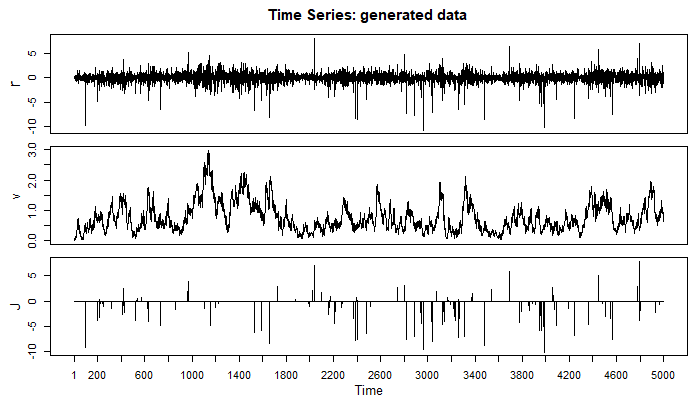}
	\caption{Simulation Study: Simulated realization (n = 5,000).}
	\label{fig1}
\end{figure}

The simulated time series consist of approximately 20 years of daily data (n = 5,000). Figure \ref{fig1} shows one realization generated from the model. All codes for the model estimation were written in R software [The R foundation for Statistical Computing, 2015], using package Rcpp, available at The Comprehensive R Archive Network (CRAN). Machine specifications are: Intel Core i7-8700K 3.70 Ghz processor, 16GB RAM, using a 64-bit Windows 10 Operating System. MCMC retained 20,000 samples after 300,000 iterations, burn-in of 60,000 and lag of 11 iterations.

\begin{figure}[H]
	\centering
	\includegraphics[width=1\linewidth]{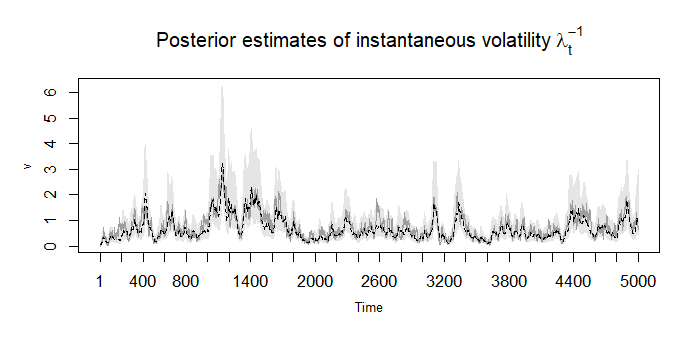}
	\caption{Simulation Study: Posterior estimates of instantaneous volatility $v_t$ for simulated data. True volatility series shown in solid gray; posterior mean estimates $\lambda_t^{-1}$ in dashed black; and 95\% credibility interval is the light gray area.}
	\label{fig2}
\end{figure}

\begin{figure}[H]
	\centering
	\includegraphics[width=1\linewidth]{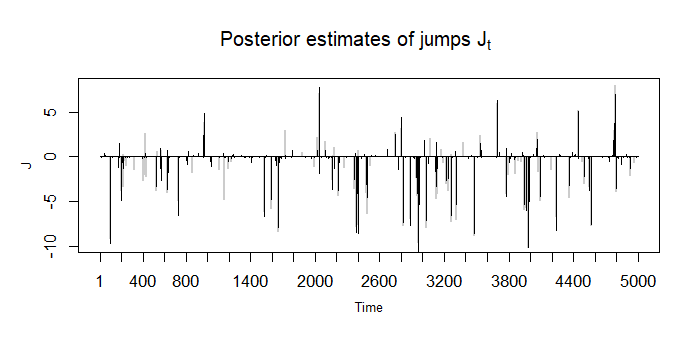}
	\caption{Simulation Study: Posterior estimates of instantaneous jumps $\mathcal{J}_t$ for simulated data. True values of the time series shown in gray; posterior mean estimates $J_t$ in black.}
	\label{fig3}
\end{figure}

Figure \ref{fig2} shows the time series of true instantaneous volatility $v_t$, together with the estimated volatility $\lambda_t^{-1}$. The NGSVJ is able to closely track the latent state. Almost every point from the true volatility is inside the 95\% credibility interval, even when the estimate mean slightly deviates from the true value.
Figure \ref{fig3} shows the time series of true instantaneous jumps $\mathcal{J}_t$, together with the estimated jumps $J_t$.
Recall that the jumps represent moments of punctual abnormal returns, caused by market's speculative movements. The NGSVJ is able to caught most of the simulated jumps, together with its magnitudes. Jump points that are not captured by the jump component of the model are propagated to volatility $\lambda^{-1}$ or heavy tail $\gamma^{-1}$ components. 

\begin{figure}[H]
	\centering
	\includegraphics[width=1\linewidth]{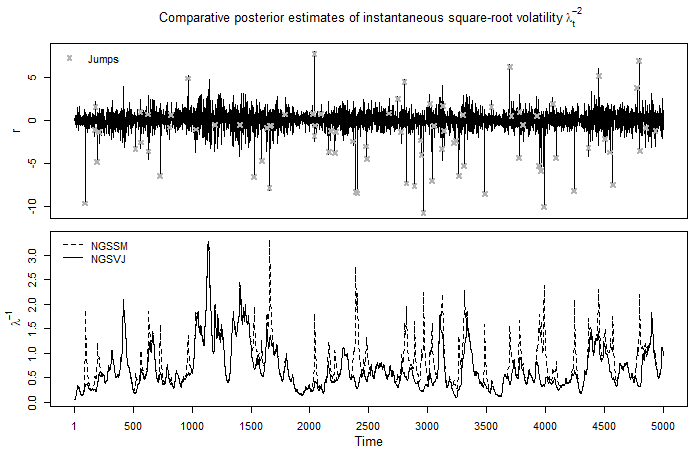}
	\caption{Simulation Study: Top graphs show posterior estimates of main instantaneous jumps $\mathcal{J}_t$ for simulated data together with simulated returns. Bottom graph shows posterior estimates for instantaneous volatility $\lambda^{-1}_t$. NGSVJ mean estimates in solid line and NGSSM (without the jump structure) mean estimates in dashed line.}
	\label{fig4}
\end{figure}

Table \ref{tab_1} presents the model estimates for each static parameter. The NGSVJ model is able to get estimates very close to true parameters. Figure \ref{fig4} compares posterior estimates of instantaneous volatility on NGSVJ and NGSSM (without jumps) models. It can be seen that the jump component absorb the impact of abnormal returns so that they do not propagate to volatility measure. Also, BIC and DIC values are smaller for NGSVJ model than NGSSM (without the jump component), indicating a better fit, despite having more parameters to estimate.

\begin{table}[H]
	\centering
	\begin{tabular}{llllllll}
		\hline
		\multicolumn{2}{c}{}&\multicolumn{3}{c}{NGSVJ}&\multicolumn{3}{c}{NGSSM}\\
		 & True & Mean & SD & RMSE & Mean & SD & RMSE\\
		\hline
		$\mu$   & 0.05   & 0.0529 & 0.0024 & 0.0037 & 0.0187 & 0.0012 & 0.0313\\
		$\rho_y$&   0.015  & 0.01544  &0.0022 & 0.0024&-&-&-\\  
		$\mu_y$&   -2.5  & -2.2799  & 0.5648& 0.6245&-&-&-\\
		$\sigma_y$    & 4 & 4.4445 & 0.4016 &0.7475&-&-&-\\
		log $\mathcal{L}$&-&\multicolumn{3}{c}{-5,783}&\multicolumn{3}{c}{-6,322}\\
		BIC&-&\multicolumn{3}{c}{11,634}&\multicolumn{3}{c}{12,669}\\
		DIC&-&\multicolumn{3}{c}{12,139}&\multicolumn{3}{c}{13,938}\\
		\hline
	\end{tabular}
	\caption{Posterior estimate of NGSVJ and NGSSM (without jumps) static parameters for simulated daily returns (n = 5,000).}
	\label{tab_1}
\end{table}

\begin{figure}[H]
	\centering
	\includegraphics[width=1\linewidth]{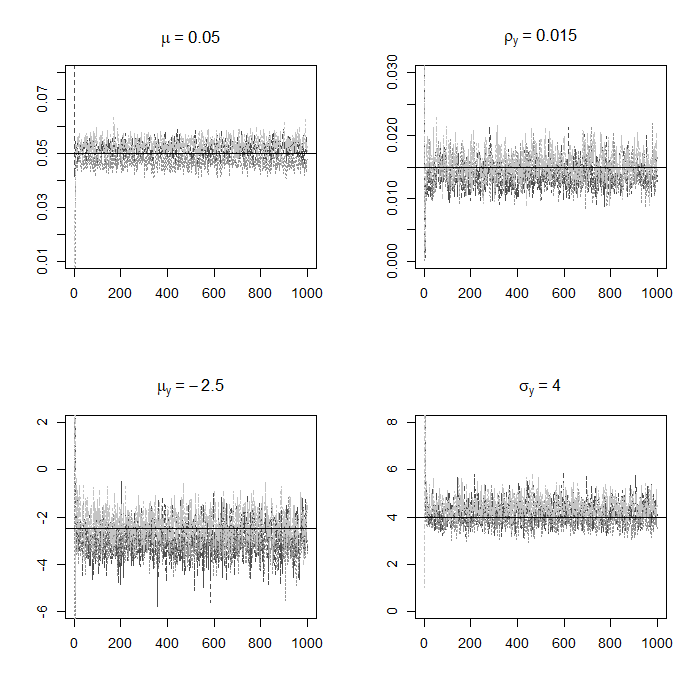}
	\caption{MCMC chain convergence: trace plot for the static parameters of the model for simulated time series using three different initial values. Each line represents a chain.}
	\label{fig5}
\end{figure}

Another advantage of the model, as can be seen on Figure \ref{fig5}, is that it rapidly achieves the convergence, due to its automatic and simple sampling structure. This allows the model to be used in on-line day-trade operations, as it will be shown on next section, to estimate market volatility and orient day-trade arbitrage strategies.

Similar results are obtained across alternative simulation scenarios, using different initial values for the parameters. As the sample size n grows very large, the dimensionality of the model raises exponentially, since there are four dynamic parameters to be estimated: volatility, $\lambda^{-1}$; mixture component, $\gamma^{-1}$; jump times, N; and jump size, $\xi$. Thus, it is limited by concerns such as computational time and available computer memory.

One way to deal with higher dimensionally data is reducing the sample size, and iterations made by the MCMC algorithm, taking advantage of the fast convergence of the chains due to the inferential procedure of the model, due to its automated characteristic via Gibbs sampler. As can be seen in Figure \ref{fig5}, convergence was achieved on the first 200 iterations, independently of the initial values for the parameters. On practical applications, we have already an idea of good initial values, based on results from other models in literature, such as Eraker \textit{et al.} (2003), Warty et al (2014),  Nakajima and Omori (2007), that can be used to achieve convergence even faster.

\section{Model applications}

In this section we show two applications of the NGSVJ model and compare it to the NGSSM proposed by Gamerman \textit{et al.} (2013) and SV model proposed by Kastner(2016), implemented on R's stochvol package available on CRAN, in order to attest its efficiency. The first deals with the estimation of S\&P500 volatility, as done by  Eraker \textit{et al.} (2003) and Warty et al (2014). The second is an application to intra-day Brent Crude future returns, to estimate volatility before the next minute return information arrives, in order to present the model as a tool for strategy making in a real decision problem, while user is operating at the market.

\subsection{S\&P500 index data}

The NGSVJ is applied to stock market index data and results will be compared to the NGSSM proposed by Gamerman \textit{et al.} (2013) and SV model proposed by Kastner(2016) for stock returns data, which do not include jumps on returns. 

The dataset contains S$\&$P 500 stock index returns from January 2, 1980, to December 31, 1999. Excluding weekends and holidays, there are 5,055 daily observations for the S$\&$P500. Table \ref{tab_d1} provides descriptive statistic for the log-returns, scaled by 100.

\begin{table}[H]
	\centering
	\begin{tabular}{ lllllll}
	\hline
	  Sample size & Mean & Variance & Skewness & Kurtosis & Min & Max\\
	 \hline	
	5,055 & 0.05205 & 0.9978 & -2.6357 &  63.0710 & -22.8997 & 8.7089 \\
	 \hline	
	\end{tabular}
	\caption{S\&P500 log-returns [$\times$100\%] descriptive statistics.}
	\label{tab_d1}
\end{table}
\begin{figure}[H]
	\centering
	\includegraphics[width=1\linewidth]{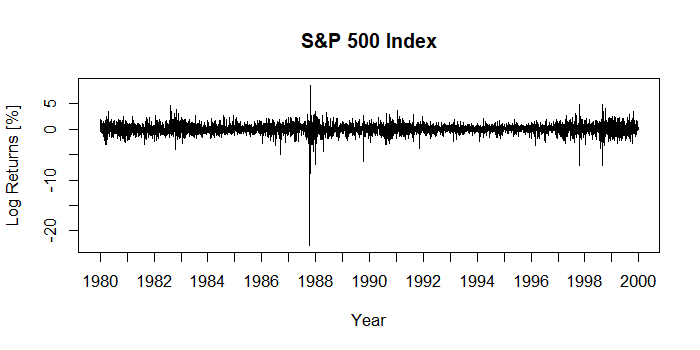}
	\caption{Log-returns of the S\&P 500 Index from January 2, 1980, to December 31, 1999
		(n = 5,055).}
	\label{fig6}
\end{figure}

\subsubsection{Parameters Specification}

For applying on stock market index data, based on sensibility analysis conclusion that suggests $\nu$ = 30, a G(15,15) prior distribution was specified for $\gamma_{t}$ in order to obtain the Student's $t_{30}$-errors for the observation and system disturbances. Recall that we avoid appealing to Metropolis steps in order to estimate $\nu$, in order to keep an automatic procedure of sampling via Gibbs sampler. The threshold, was fixed at $\alpha=0.7$ and the discount factor $\omega$ fixed at 0.9. For the mean components, $\mu$ and $\mu_{y}$, N(0,100) priors were specified and for $\sigma^{2}_{y}$ a InverseGamma(0.1,0.1) prior. Also, $a_{0}=0.1$ and $b_{0}=0.1$, as suggested in West $et$ $al.$ (1987,p. 333), cited by Gamerman \textit{et al.} (2013). A Beta(2,40) prior distribution was specified to $\rho_{y}$, as in Eraker \textit{et al.} (2003). For the NGSSM model the same specifications were made, except that this model dos not include the jump components.

The results were obtained with a 300,000 iteration chain, burn-in of 60,000 observations, with a lag of 11 observations, resulting in 20,000 samples. MCMC chains convergence was verified through graphic methods. All programing was done in  the R software (The R foundation for Statistical Computing, 2015), using Rcpp package.

\subsubsection{Results}

Table \ref{tab_3} shows model estimates for each of the static parameters. For NGSVJ, as observed by Eraker \textit{et al.} (2003), it is possible to see that jumps in returns are infrequent, since jump probability $\rho_y$ is small, but have a large magnitude, as can be seen by the magnitude of jump size mean, $\mu_y$. BIC and DIC criteria favors NGSVJ over NGSSM and SV models, which indicates the former has better fit to the data. Computational time for NGSVJ is close to SV model, which makes it competitive, since it includes the jump structure that advantage of the automatic inference procedure to boost it speed. The inclusion of jumps makes NGSVJ 55.6\% slower than NGSSM model, but it has a better fit and a more precise estimate for volatility, since abnormal returns are captured by jump component.

\begin{figure}[H]
	\centering
	\includegraphics[width=1\linewidth]{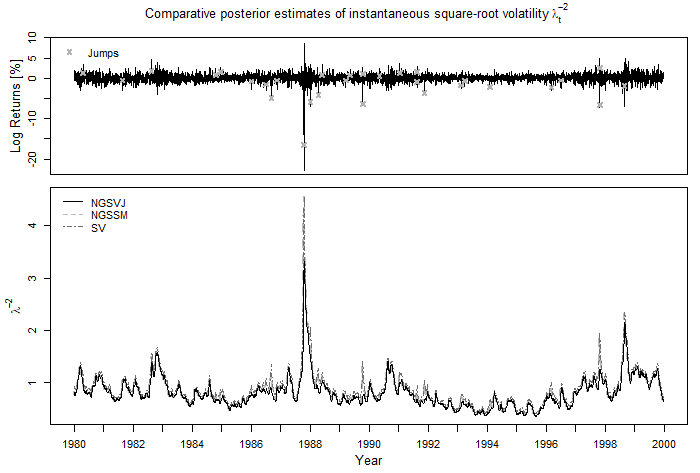}
	\caption{Top graph shows posterior estimates of main instantaneous jumps $J_t$ for S\&P500 together with  log-returns [$\times$100\%]. Bottom graph shows posterior estimates of instantaneous square-root of volatility $\lambda_t^{-2}$ for S\&P 500 Index data. NGSVJ mean estimates in solid line; NGSSM mean estimates in dashed line; and SV mean estimates in dotdashed line.}
	\label{fig7}
\end{figure}

\begin{table}[H]
	\centering
	\begin{tabular}{lllllll}
		\hline
		&\multicolumn{2}{c}{NGSVJ}&\multicolumn{2}{c}{NGSSM}&\multicolumn{2}{c}{SV} \\
		& Mean & SD & Mean & SD & Mean & SD\\
		\hline
		$\mu$   & 0.0616   & 0.0020 &0.0522 & 0.0012&0&- \\
		$\rho_y$&  0.0042  & 0.0012  &-& - &-& -\\  
		$\mu_y$&   -2.4598  & 1.4302  &-& - &-& -\\
		$\sigma_y$    & 5.2793 & 1.2598 & - &- &-& -\\
		log $\mathcal{L}$  &\multicolumn{2}{c}{-5,972} &\multicolumn{2}{c}{-6,086}&\multicolumn{2}{c}{-8,468}\\
		BIC &\multicolumn{2}{c}{12,012} &\multicolumn{2}{c}{12,197}&\multicolumn{2}{c}{12,206}\\
		DIC &\multicolumn{2}{c}{12,338} &\multicolumn{2}{c}{12,763}&\multicolumn{2}{c}{21,719}\\
		Comp. time  &\multicolumn{2}{c}{738} &\multicolumn{2}{c}{474} &\multicolumn{2}{c}{753}\\
		\hline
	\end{tabular}
	\caption{Posterior inference of static parameters for NGSVJ, NGSSM and SV models for S\&P500 daily returns. Computational time is given in seconds.}
	\label{tab_3}
\end{table}
Figure \ref{fig7} shows the posterior mean estimates of instantaneous square root volatility $\lambda^{-2}$ for NGSVJ, NGSSM and SV models. As expected, the NGSVJ model estimates a lower magnitude volatility measure, since part of the log-returns variation is absorbed by the jump component, so they do not propagate to volatility as an increase on risk.

\begin{figure}[H]
	\centering
	\includegraphics[width=1\linewidth]{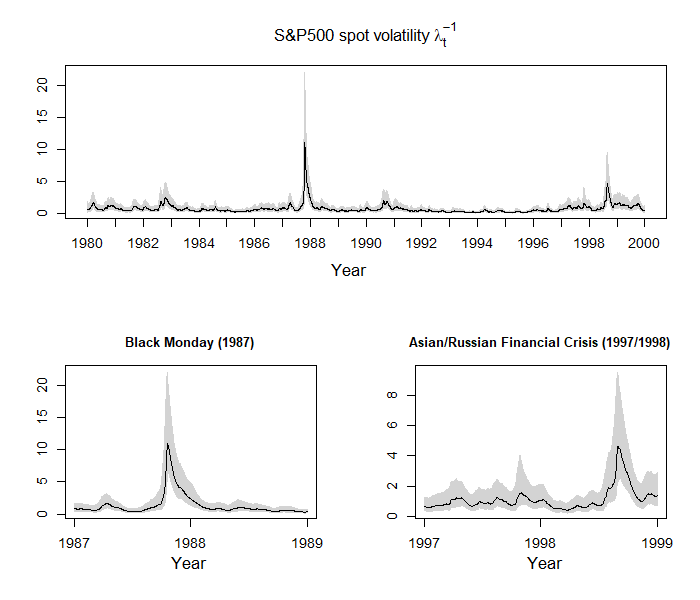}
	\caption{Posterior estimates of instantaneous volatility $\lambda_t^{-1}$ for S\&P 500 Index data. NGSVJ mean estimates in solid line; and 95\% credibility interval is the gray area.}
	\label{fig8}
\end{figure}

Figure \ref{fig8} shows the posterior mean estimates of instantaneous volatility $\lambda^{-1}$ for NGSVJ with the 95\% credibility interval. Also, a zoom into two specific moments known by market crisis: the Black Monday (1987) and the Asian/Russian financial crisis (1997,1998). Continuous line represents the posterior mean and the gray area indicates the 95\% percentile credibility interval for spot volatility. Results are consistent with Eraker \textit{et al.} (2003) findings, with volatility peaks occurring at the same time.

\begin{figure}[H]
	\centering
	\includegraphics[width=1\linewidth]{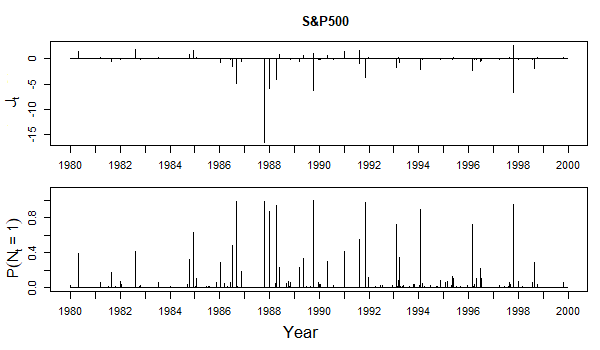}
	\caption{Posterior estimates of jump times, N, and jump sizes, J, for S\&P 500 Index data from NGSVJ model.}
	\label{fig9}
\end{figure}

Figure \ref{fig9} provides jump sizes and probabilities for each observation. Before moments of higher volatility, it is possible to observe an increase in jump sizes and probabilities, when compared to periods with lower volatility, thus, evidencing that such moments are preceded by speculative movements, captured in the model as jumps. 

\subsection{Intra-day returns data}

The NGSVJ model is applied to intra-day returns of Brent Crude Futures, so that a market agent can use this information to choose strategies on day-trade operations. In order to be used during market operations, the model must be able to deliver reliable and fast results. The proposed model performance will be compared to NGSSM and SV models.

The dataset contains Brent Crude futures, ICE:BRN, log returns from August 13, 2018 to August 17, 2018, in a total of 6,241 minute observations. Table below provides summary statistics for the log returns scaled by 100.

\begin{table}[H]
	\centering
	\begin{tabular}{ lllllll}
		\hline
		Sample size & Mean & Variance & Skewness & Kurtosis & Min & Max\\
		\hline	
		6,241 & -0.00026 & 0.00204 & -1.6751 &  42.3833 & -0.95748 & 0.39991 \\
		\hline	
	\end{tabular}
	\caption{Descriptive statistics of ICE:BRN log-returns [$\times$100\%]}
	\label{tab_4}
\end{table}

\subsubsection{Parameters Specifications}

The NGSVJ setup is same specifications defined on section 4.1.1 where applied. The results were obtained with a 10,000 iteration chain, burn-in of 6,000 observations with a lag of 2 samples, resulting in 1,667 samples. MCMC chains convergence was verified through graphic methods. This specifications takes advantage of the fast convergence so that the model can give results in about 30 seconds, so that an strategy decision can be taken before the next minutely observation arrives.

\subsubsection{Results}

Table \ref{tab_5} shows model estimates for each of the static parameters. BIC and DIC criteria strongly favors NGSVJ and NGSSM and SV models, since chain convergence is not achieved for all parameters on the latter model. Both NGSVJ and NGSSM models are able to deliver reliable results in less then one minute, before the next information arrives, but the former has the advantage of including jumps in returns, thus, providing a better fit to the data and a more precise volatility estimate.

\begin{table}[H]
	\centering
	\begin{tabular}{lllllll}
		\hline
		&\multicolumn{2}{c}{NGSVJ}&\multicolumn{2}{c}{NGSSM}&\multicolumn{2}{c}{SV} \\
		& Mean & SD & Mean & SD & Mean & SD\\
		\hline
		$\mu$   & 0.000106   & 0.000101 &-0.000099 & 0.000001&0&- \\
		$\rho_y$& 0.008716  & 0.001532  &-& - &-& -\\  
		$\mu_y$&  -0.023048  & 0.033438  &-& - &-& -\\
		$\sigma_y$    & 0.049826 & 0.011876 & - &- &-& -\\
		log $\mathcal{L}$  &\multicolumn{2}{c}{12737} &\multicolumn{2}{c}{12407}&\multicolumn{2}{c}{4235}\\
		BIC &\multicolumn{2}{c}{-25405} &\multicolumn{2}{c}{-24787}&\multicolumn{2}{c}{-8435}\\
		DIC &\multicolumn{2}{c}{-24919} &\multicolumn{2}{c}{ -23747}&\multicolumn{2}{c}{-8409}\\
		Comp. time  &\multicolumn{2}{c}{31.83} &\multicolumn{2}{c}{20.60} &\multicolumn{2}{c}{28.64}\\
		\hline
	\end{tabular}
	\caption{Posterior inference of static parameters for NGSVJ, NGSSM and SV models for ICE:BRN minutely returns. Computational time is given in seconds.}
	\label{tab_5}
\end{table}

\begin{figure}[H]
	\centering
	\includegraphics[width=1\linewidth]{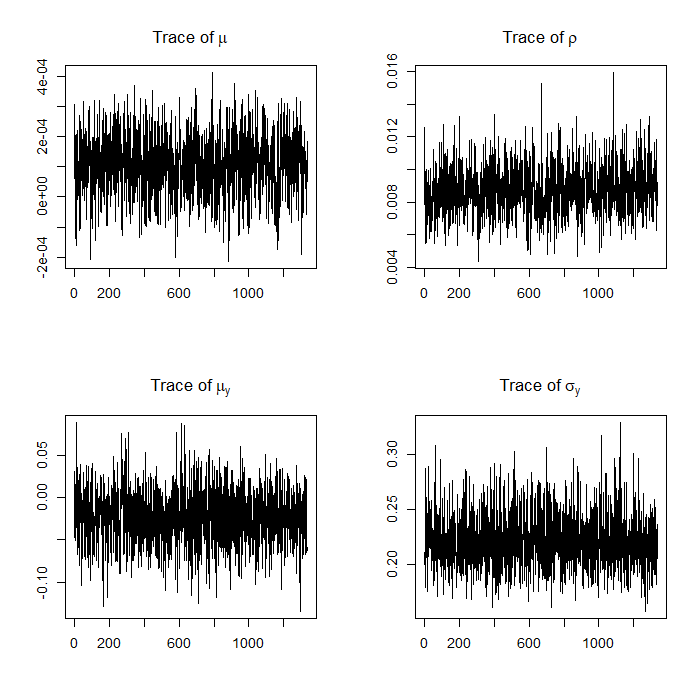}
	\caption{MCMC chain convergence: trace plot for static parameters for ICE:BRN time series for NGSVJ model, after burn-in and lags. Convergence is achieved for all static parameters.}
	\label{fig10}
\end{figure}

Figure \ref{fig10} shows trace plot for static parameters for ICE:BRN time series for NGSVJ model, after burn-in and lags. As seen on simulation study the model is able to rapidly achieve convergence de to its automatic and simple sampling structure.  This property makes NGSVJ both fast enough to provide results in a minutely time frame and reliable, so that the user can trust on this results in order to take strategy decisions.

\begin{figure}[H]
	\centering
	\includegraphics[width=1\linewidth]{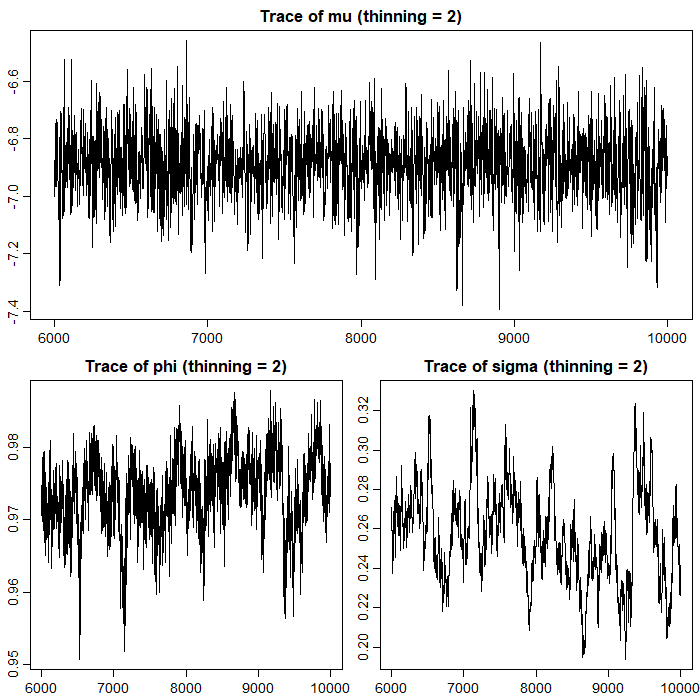}
	\caption{MCMC chain convergence: trace plot for static parameters for ICE:BRN time series for SV model, provided by package stochvol, after burn-in. Convergence is not achieved for all static parameters.}
	\label{fig11}
\end{figure}

Figure \ref{fig11} shows trace plot for static parameters for ICE:BRN time series for SV model, provided by package stochvol, after burn-in. Despite being fast enough to bring results on the required time frame, the SV model was not able to achieve convergence for all parameters. It can be seen that phi and sigma parameters have still not converged after six thousand iterations, thus, such results are not reliable to be used for taking strategy decisions.

\begin{figure}[H]
	\centering
	\includegraphics[width=1\linewidth]{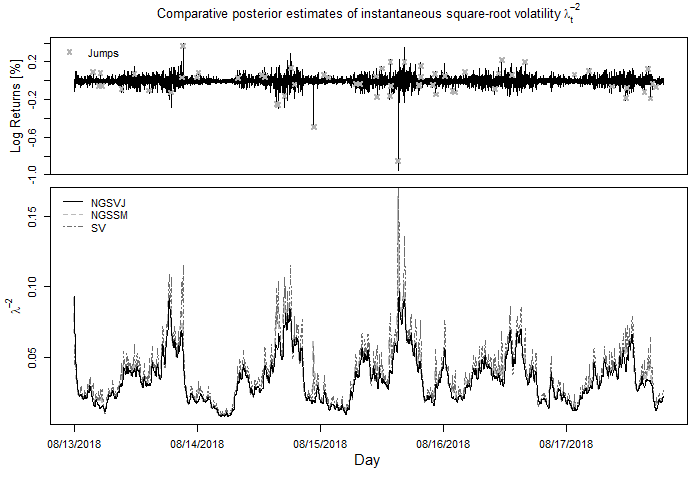}
	\caption{Top graph shows posterior estimates of main instantaneous jumps $J_t$ for ICE:BRN together with  log-returns [$\times$100\%]. Bottom graph shows posterior estimates of instantaneous square-root of volatility $\lambda_t^{-2}$ for ICE:BRN data. NGSVJ mean estimates in solid line; NGSSM mean estimates in dashed line; and SV mean estimates in dotdashed line.}
	\label{fig12}
\end{figure}

Figure \ref{fig12} shows the posterior mean estimates of instantaneous square root volatility $\lambda^{-2}$ for NGSVJ, NGSSM and SV models. As observed before, the NGSVJ model estimates a lower magnitude volatility measure, since part of the log-returns variation is absorbed by the jump component, so they do not propagate to volatility as an increase on risk.

\begin{figure}[H]
	\centering
	\includegraphics[width=1\linewidth]{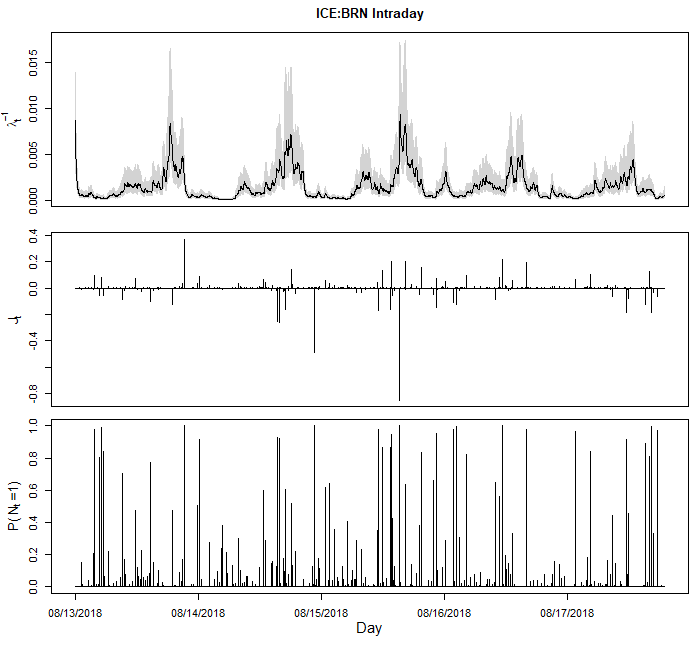}
	\caption{Top graph shows posterior estimates of instantaneous volatility $\lambda_t^{-1}$ for ICE:BRN data. NGSVJ mean estimates in solid line; and 95\% credibility interval is the gray area;
	Middle graph shows posterior mean estimates for jump sizes, $J_t$; and posterior estimates for jump times $N_t$ are shown on the bottom graph.}
	\label{fig13}
\end{figure}

Figure \ref{fig13} shows the posterior mean estimates of instantaneous volatility $\lambda^{-1}$ for NGSVJ with the 95\% credibility interval, jump sizes and probabilities for each observation. Continuous line represents the posterior mean and the gray area indicates the 95\% percentile credibility interval for spot volatility.

\section{Discussion} 

The NGSVJ model was able to capture speculative movements in the market through the jump components and detect periods with increased market risk through the volatility component. Results obtained by applying the model to S\&P 500 return series are consistent with Eraker \textit{et al.} (2003) findings, since volatility peaks occur at same times. Historical events of known volatility effects on financial markets, such as Black Monday and Asian/Russian Financial Crisis, are detected by the model precisely. Also, the inclusion of jumps improve the model fit when compared to NGSSM and SV models.

The most notable advantage of using NGSVJ is its computational simplicity and the structure of the NGSSM, and grants an automatic sampling process for parameters, that allows to sample the volatility in block via Gibbs sampler. This structure allows model parameters to achieve convergence with less MCMC iterations, providing fast and reliable estimates and allowing the model to be used on practical situations, such as taking decisions on intra-day operations arbitrage strategies.

Using Gibbs Sampler to draw a sample from conditional posterior distributions is computationally cheaper than recurring to Metropolis based algorithms. Since Metropolis is an accept-reject algorithm, it can take several steps until a full representative sample is obtained in order to make suitable statistical inferences, unless a very good proposal distribution is given to the algorithm. On using Gibbs Sampler, a representative sample can potentially be obtained in less steps. This is specially relevant when dealing with small time frames, as in intra-day operations.

As the model is built over the Dynamic linear model with scale mixtures proposed by Gamerman \textit{et al.}(2013) an exact sample of volatility parameter $\lambda^{-1}_{t}$ is drawn. The three main advantages that comes from this process are: there is no need to make approximations in order to estimate volatility; the volatility is sampled in blocks from the proposed model; and there is no need to appeal to Metropolis based algorithms. 

Another advantage is the model flexibility. It can be adapted to include jumps, covariates, heavy tails and different distributions can be adopted based on the mixture component used. In this work, Student-t distribution was obtained through a Gamma mixed component, but other distributions can also be used to give satisfactory results. Also, it was shown that the inclusion of jumps raises the performance of the model, when compared to the NGSSM model.

The inclusion of jumps in the model reduces substantially the volatility estimate. This would have several impacts on risk analysis, since less volatility indicates less risk, in other words, knowing that some event was a jump, or tail event, and not a recurrent market event would mean that such asset is still a safe bet. Usually an increase of jump frequency is observed near the occurrences of market anomalies, such as Black Monday and the Asian and Russian financial crisis, which may indicate that it can also be used for predicting near future increases on market risk.

For day-to-day operations, the automatic model is effective and can be used in order to estimate market volatility, which can be used for options pricing, VaR calculations, measuring market regimes, etc. 

For future works is intended to extend the model to the multivariate case, where, instead of analysing one asset individually, an asset portfolio risk is analysed as a whole. Another extension is the inclusion of jumps in volatility, as suggested by Eraker \textit{et al.} (2003), without having to appeal to Metropolis based algorithms, in order to keep the inferential procedure of the model fast and accessible. Some other possibilities include working with a skew heavy-tailed distribution, as in Nakajima and Omori (2007) but still maintaining the block sampling, so that it allows to capture the leverage effect and correlations between mean and volatility.


\clearpage
\appendix
\section{Sampling from $\lambda_{t}$}
$$ $$

This appendix shows how the posterior sample from  $\lambda$ is drawn [Gamerman \textit{et al.}, 2013].

The joint distribution of $p(\lambda|Y_{n},\varphi)$ has density
$$ p(\lambda|\varphi,Y_{n}) = p(\lambda_{n}|\varphi,Y_{n})\prod_{t=1}^{n-1}p(\lambda_{t}|\lambda_{t+1},\varphi,Y_{t})p(\varphi|Y_{n}) $$
where the distribution of $(\lambda_{t}|\lambda_{t+1},\varphi,Y_{t})$ is given by
$$\lambda_{t}-\omega\lambda_{t+1}|\lambda_{t+1},\varphi,Y_{t} \sim Gamma((1-\omega)a_{t},b_{t}), \forall t \geq 0, $$
where $a_{t}$ and $b_{t}$ are the filtering parameters. The on-line or updated distribution $\lambda_{t}|\varphi, Y_{t}$, where  $\varphi = (\theta_{t},J^{y}_{t}, \gamma_{t},\omega)$, is given by:

\begin{eqnarray*}
	p(\lambda_{t}| \varphi, Y_{t-1})&\sim&Gamma(\omega a_{t-1},\omega b_{t-1}) \\
	p(\lambda_{t}|\varphi,Y_{t})&\propto&(\lambda_{t})^{\omega a_{t-1}-1}e^{-\lambda_{t}\omega b_{t-1}}\left(\frac{1}{\gamma_{t}^{-1}\lambda_{t}^{-1}}\right)^{\frac{1}{2}}\exp\left(-\frac{(Y_{t}-F_{t}\theta_{t}-J_{t})^{2}}{2\gamma_{t}^{-1}\lambda_{t}^{-1}}\right) \\
	p(\lambda_{t}|\varphi,Y_{t})&\propto&(\lambda_{t})^{\omega a_{t-1}+\frac{1}{2}-1}\exp\left(-\lambda_{t} \left[ \omega b_{t-1}+\gamma_{t}\frac{(Y_{t}-F_{t}\theta_{t}-J_{t})^{2}}{2}\right]\right) \\
	p(\lambda_{t}|\varphi,Y_{t})&\sim&Gamma\left(\omega a_{t-1}+\frac{1}{2}, \omega b_{t-1}+\gamma_{t}\frac{(Y_{t}-F_{t}\theta_{t}-J_{t})^{2}}{2}\right) 
\end{eqnarray*}

Based on that Theorem and a sample of $p(\varphi|Y_{n})$, an exact sample of the joint distribution $(\lambda|Y_{n},\varphi)$ can be obtained following the algorithm:
\begin{enumerate}
	\item set t = n and sample $p(\lambda_{n}|\varphi,Y_{n})$;
	\item set t = t-1 and sample $p(\lambda_{t}|\lambda_{t+1},\varphi,Y_{t})$;
	\item if t $>$ 1, go back to step 2; otherwise, the sample of $(\lambda_{1},...,\lambda_{n}|\varphi,Y_{n})$ is complete.
\end{enumerate}

This procedure allows the implementation of the algorithm, described in section 2.2, step \textit{iv} and enables to obtain an exact sample from the smoothed distribution of the states conditioned on other parameters.


\end{document}